\documentclass[sigconf, natbib=true]{acmart}
\usepackage{subfig}
\usepackage{multirow}
\usepackage{xspace}
\usepackage{amsmath}
\usepackage[normalem]{ulem}
\usepackage{algorithm}
\usepackage{algpseudocode}
\newcommand{\modelname}{CL4FedRec\xspace} 

\newcommand{\rmodelname}{rCL4FedRec\xspace}

\AtBeginDocument{%
  }

\begin{document}

\title{Robust Federated Contrastive Recommender System against Model Poisoning Attack}

\author{Wei Yuan}
\affiliation{%
  \institution{The University of Queensland}
  \city{Brisbane}
  \state{QLD}
  \country{Australia}
}
\email{w.yuan@uq.edu.au}

\author{Chaoqun Yang}
\affiliation{%
\institution{Griffith University}
  \city{Gold Coast}
  \state{QLD}
  \country{Australia}
}
\email{chaoqun.yang@griffith.edu.au}

\author{Liang Qu}
\affiliation{%
  \institution{The University of Queensland}
  \city{Brisbane}
  \state{QLD}
  \country{Australia}
}
\email{l.qu1@uq.edu.au}

\author{Guanhua Ye}
\affiliation{%
 \institution{Deep Neural Computing Company Limited}
 \city{Shenzhen}
 \country{China}}
\email{rex.ye@dncc.tech}

\author{Quoc Viet Hung Nguyen}
\affiliation{%
  \institution{Griffith University}
  \city{Gold Coast}
  \state{QLD}
  \country{Australia}
}
\email{henry.nguyen@griffith.edu.au}

\author{Hongzhi Yin}\authornote{Corresponding author.}
\affiliation{%
  \institution{The University of Queensland}
  \city{Brisbane}
  \state{QLD}
  \country{Australia}
}
\email{db.hongzhi@gmail.com}

\renewcommand{\shortauthors}{Yuan et al.}

\begin{abstract}
Federated Recommender Systems (FedRecs) have garnered increasing attention recently, thanks to their privacy-preserving benefits. However, the decentralized and open characteristics of current FedRecs present two dilemmas. First, the performance of FedRecs is compromised due to highly sparse on-device data for each client. Second, the system's robustness is undermined by the vulnerability to model poisoning attacks launched by malicious users. In this paper, we introduce a novel contrastive learning framework designed to fully leverage the client's sparse data through embedding augmentation, referred to as \modelname. Unlike previous contrastive learning approaches in FedRecs that necessitate clients to share their private parameters, our \modelname aligns with the basic FedRec learning protocol, ensuring compatibility with most existing FedRec implementations. We then evaluate the robustness of FedRecs equipped with \modelname by subjecting it to several state-of-the-art model poisoning attacks. Surprisingly, our observations reveal that contrastive learning tends to exacerbate the vulnerability of FedRecs to these attacks. This is attributed to the enhanced embedding uniformity, making the polluted target item embedding easily proximate to popular items. Based on this insight, we propose an enhanced and robust version of \modelname (\rmodelname) by introducing a regularizer to maintain the distance among item embeddings with different popularity levels. Extensive experiments conducted on four commonly used recommendation datasets demonstrate that \rmodelname significantly enhances both the model's performance and the robustness of FedRecs.
\end{abstract}






\maketitle

\section{Introduction}
Recommender systems have evolved into an indispensable component of numerous online services (e.g., social media~\cite{yin2015dynamic,wang2017location}, e-commerce~\cite{wei2007survey}, and online news~\cite{wu2020mind}) to assist users in identifying the potential interests among massive information.
Traditionally, recommender systems have been built by leveraging user personal data gathered and stored in a centralized server~\cite{zhang2019deep}. Nevertheless, in light of the increasing privacy concerns and the advent of stringent privacy protection regulations, such as the General Data Protection Regulation (GDPR\footnote{\url{https://gdpr-info.eu/}}) and the California Consumer Privacy Act (CCPA\footnote{\url{https://oag.ca.gov/privacy/ccpa}}), conventional centralized recommender systems encounter the imminent risks of leaking sensitive user information and running afoul of these regulatory frameworks~\cite{lam2006you}.

To address privacy concerns, federated learning~\cite{kairouz2021advances}, a privacy-preserving training paradigm, has been embraced in recommender systems, giving rise to federated recommender systems (FedRecs)~\cite{yang2020federated}.
Within FedRecs, the recommender model is partitioned into public and private parameters. 
Private parameters, such as user embeddings, are locally maintained on the user side, while public parameters, like item embeddings, are exchanged between users and the central server to facilitate collaborative learning. 
Throughout the training process, users/clients\footnote{In this paper, one client represents one user; therefore, the concepts of ``user'' and ``client'' are equivalent and can be interchangeably used.} train their local recommender models using personal data and subsequently upload the public parameter updates to a central server for aggregation. 
This design makes users' private data undisclosed to other participants. 
Owing to the privacy-preserving nature, FedRecs have garnered growing attention and demonstrated notable achievements recently~\cite{ammad2019federated,chai2020secure,wu2022federated,sun2022survey}.

While the decentralized and open characteristics of FedRecs offer a certain level of privacy guarantee, they simultaneously face two challenges.
Firstly, each user's own data is extremely sparse. Training on these sparse data dramatically increases the training difficulty of FedRecs and even compromises the model performance~\cite{wang2021fast}.
Secondly, since all participants can directly upload model updates to modify the recommender model, FedRecs become susceptible to model poisoning attacks, in which malicious users send poisoned updates to manipulate the recommender system to achieve adversarial goals (e.g., promote or demote target items)~\cite{zhang2022pipattack}.

Contrastive learning is one of the most popular research lines to tackle the data sparsity problem in centralized recommender systems~\cite{yu2023self}. The core idea of these contrastive learning methods is to design multiple views for diverse data or learned representations, followed by the maximization of agreement between these distinct views~\cite{jing2023contrastive}.
Unfortunately, applying contrastive learning in FedRecs is nontrivial since clients can only leverage their own resources.
Specifically, clients cannot obtain negative samples for user embeddings as they can only access their own private parameters; meanwhile, constructing high-quality item views on local data is also difficult.
To solve this problem, some works~\cite{wu2022fedcl,luo2023perfedrec++} break the basic FedRec learning protocol by permitting participants to utilize others' private user embeddings, raising privacy concerns.
~\cite{yu2023untargeted} overlooks the role of user embedding and exclusively applies contrastive learning to item embeddings by coarsely utilizing interacted items as positive samples and non-interacted items as negative samples, which is less effective for improving model performance.
Consequently, implementing an effective federated contrastive recommender system without compromising privacy is still under-explored.

In this paper, we propose a \underline{c}ontrastive \underline{l}earning framework tailored \underline{for} \underline{fed}erated \underline{rec}ommendation, namely \modelname.
In this method, to facilitate user view construction, we generate a group of synthetic users on the central server, serving as negative users for all real participants.
Then, inspired by~\cite{yu2022graph}, we incorporate lightweight uniform noises to user embeddings to create positive pairs. 
Based on the augmented user representations, we enhance two distinct views of item embeddings by approximately optimizing them on local data.
We do not directly add the uniform noises~\cite{yu2022graph} to item embedding as we empirically find it cannot achieve satisfied recommendation performance.
This may stem from the relatively small datasets within each FedRec client, preventing the model from learning good item representations from the augmentation based on uniform noises.
It is essential to highlight that \modelname functions as an auxiliary task that can be jointly optimized with the recommendation task on the client side and is consistent with the basic FedRec learning protocol. Therefore, it is compatible with most existing FedRecs.

After that, we empirically assess the robustness of FedRecs equipped with \modelname using several state-of-the-art model poisoning attacks. 
Unexpectedly, we find that contrastive learning intensifies the susceptibility of FedRecs to these poisoning attacks.
We then conduct extensive analysis, attributing this phenomenon to the uniformity of the representation distribution enforced by the contrastive learning task.
Specifically, in a more uniform embedding space, adversaries can easily modify target item embeddings to mimic popular ones.
In light of this, we propose a robust \modelname (\rmodelname) by adding a regularizer in the central server.
The regularizer is designed to maintain the distance among item embeddings with various popularity levels, thwarting malicious users from easily boosting the exposure rate of target items by aligning them with arbitrary popular items. To demonstrate the effectiveness of our proposed methods, we conduct extensive experiments on four recommendation datasets (MovieLens-1M~\cite{harper2015movielens}, Amazon-Phone, Amazon-Video~\cite{mcauley2015image}, and QB-Article~\cite{yuan2022tenrec}) with several commonly used FedRecs.
The experimental results showcase that \rmodelname can improve the recommendation effectiveness and make FedRecs more robust to model poisoning attacks.

To sum up, the main contributions of this paper are fourfold:
\begin{itemize}
  \item We propose the first contrastive learning framework tailored for federated recommendation (\modelname) without privacy compromise.
  \item  Through empirical analysis, we find that contrastive learning significantly reduces FedRecs' resistance to model poisoning attacks because of the improved uniformity of embeddings. To the best of our knowledge, this work is the first to reveal such an interesting phenomenon in federated recommendation.
  \item We further propose a robust version of \modelname, named \rmodelname, by incorporating a popularity-based contrastive regularizer. It is noteworthy that, unlike traditional defense methods, our regularizer does not compromise but enhances the recommendation effectiveness.
  \item Extensive experiments show that our proposed contrastive learning frameworks for FedRecs significantly improve recommendation effectiveness and robustness to poisoning attacks.
\end{itemize}

\section{Related Work}
\subsection{Federated Recommender System}
The privacy-preserving ability has made FedRecs receive remarkable focus in recent years~\cite{sun2022survey,yang2020federated,long2023decentralized,qu2023semi,long2024physical,qu2024towards,yin2024device}.
Ammand et al.~\cite{ammad2019federated} propose the first federated recommender system with collaborative filtering models. 
Building upon this foundational framework, numerous extended works have emerged.
For example,~\cite{chai2020secure} introduces a user-level distributed matrix factorization FedRec framework, incorporating homomorphic encryption to enhance user privacy protection.
~\cite{wu2022federated} explores the utilization of graph neural network~\cite{scarselli2008graph}, assuming the availability of a trusted third-party server.
~\cite{qi2020privacy,liu2022federated,guo2021prefer,liang2021fedrec++} apply FedRecs to different recommendation scenarios. 

\textbf{Robustness of FedRecs.} 
With notable achievements, many researchers have redirected their focus to verifying the security of FedRecs by investigating poisoning attacks.
Generally, based on adversarial goals, these attacks can be categorized into targeted attacks~\cite{zhang2022pipattack,rong2022fedrecattack,rong2022poisoning,yuan2023manipulating,yuan2023manipulating1} and untargeted attacks~\cite{yu2023untargeted,wu2022fedattack}.
The former aims to recommend specific items to most users, while the latter endeavors to compromise recommendation performance. 
Zhang et al.~\cite{zhang2022pipattack} present Pipattack, the first model poisoning attack capable of adversarially promoting items in FedRecs. 
However, Pipattack necessitates a substantial number of malicious users. 
FedRecAttack~\cite{rong2022fedrecattack} reduces the requirement for malicious users but relies on the strong assumption that adversaries can observe a proportion of user interaction data.
~\cite{rong2022poisoning} and~\cite{yuan2023manipulating} are two recently released state-of-the-art model poisoning attacks that can promote items to other participants with much fewer costs. 
In this paper, we will leverage these two attacks to assess the robustness of FedRecs.

\textbf{Contrastive Learning in FedRecs.}
While contrastive learning has demonstrated potent representation learning capabilities in centralized recommender systems~\cite{jing2023contrastive,zhang2021double}, its application in FedRecs remains relatively unexplored.
~\cite{wu2022fedcl,luo2023perfedrec++,luo2022dual} incorporate contrastive learning in FedRecs, yet they deviate from the FedRec learning protocol to grant participants access to others' private parameters, posing potential privacy concerns.
~\cite{yu2023untargeted} introduces a contrastive learning approach named UNION in FedRecs, however, the primary focus of UNION is to detect untargeted attacks rather than improve model performance.

\subsection{Contrastive Learning in Recommender System}\label{sec_related_cl_in_rs}
The comprehensive introduction of contrastive learning-based recommender can refer to some recent surveys~\cite{yu2023self,jing2023contrastive}.
Essentially, the core concept of contrastive learning is to maximize agreement between different views. 
Therefore, unless multiple views naturally exist in some cases~\cite{zhou2020s3,wei2021contrastive}, the primary focus of research in contrastive learning-based recommender systems is the construction of high-quality views, known as augmentation.
Augmentation can be categorized into two research lines: data-based augmentation and model-based augmentation.
The data-based augmentation~\cite{xia2021self,yu2021self,wu2021self,xie2022contrastive} primarily involves perturbations based on specific data attributes.
For instance, edge and node dropout are widely employed for graph data~\cite{wu2021self}, while sequence shuffling and item masking serve as common augmentation operators for sequential data~\cite{xie2022contrastive}.
Nonetheless, in FedRecs, where each client has limited data, often in the form of a one-hop bipartite graph or a very short sequence, these data augmentation methods tend to be infeasible.
The model-based augmentation aims to directly perturb the recommendation model to create distinguishable views.
One notable work is~\cite{yu2022graph}, which suggests that by simply applying uniform noises, contrastive learning can achieve comparable or even superior performance to data augmentation. 
Unfortunately, our subsequent experiments reveal that this simplistic augmentation method fails to deliver good performance in FedRecs, possibly due to the exceedingly limited data in clients, which cannot support meaningful model learning from random noise perturbation.

\section{Preliminaries}
In this section, we first introduce the primary settings of the general federated recommender systems.
Then, we briefly introduce the model poisoning threats in the federated recommendation.

\subsection{General Federated Recommender System Framework}\label{sec_prim_fedrec}
Following most FedRec robustness works~\cite{zhang2022pipattack,rong2022fedrecattack,rong2022poisoning,yuan2023manipulating}, our research is based on the most general federated recommendation framework described as follows.

Let $\mathcal{U}$ and $\mathcal{V}$ represent\footnote{In this paper, bold lowercase (e.g., $\mathbf{a}$) signifies vectors, bold uppercase denotes matrices (e.g., $\mathbf{A}$), and squiggly uppercase represents functions or sets (e.g., $\mathcal{A}$). } the sets of all users (clients) and items in FedRecs.
Each user/client $u_{i}$ possesses a local training dataset $\mathcal{D}_{i}$ containing a few user-item interaction records $(u_{i}, v_{j}, r_{ij})$.
$r_{ij}=1$ indicates that the user has interacted with item $v_{j}$, while $r_{ij}=0$ implies $v_{j}$ is a negative sample.
Note that due to the data sparsity in recommender systems~\cite{yin2020overcoming}, the size of each local dataset $\left|\mathcal{D}_{i}\right|$ is usually small.
The goal of FedRecs is to train a recommender model based on distributed datasets $\{\mathcal{D}_{i}\}_{u_{i}\in\mathcal{U}}$ that can predict users' preference scores $\hat{r}_{ij}$ for non-interacted items and make top-K recommendations based on these scores.

\textbf{General FedRec Learning Protocol.} 
Generally, FedRecs collaboratively learn a recommender model under the following learning protocol.
A central server functions as a coordinator, determining which clients will participate in the current training round. The recommender model's parameters are categorized into two groups: public parameters and private parameters.
The private parameters are user embeddings $\mathbf{U}$, encompassing users' sensitive attributes. 
Consequently, these parameters are locally managed by the respective users.
Public parameters include item embeddings $\mathbf{V}$ and other model parameters $\mathbf{\Theta}$, which clients collaboratively update.
In the initial stage, the central server initializes the public parameters while the clients initialize their private parameters. 
Subsequently, the recommender model undergoes training by repetitively executing the following steps.
Firstly, the central server randomly selects a subset of users to contribute to the current round's training and disperses public parameters $\mathbf{V}^{t-1}$ and $\mathbf{\Theta}^{t-1}$ to these clients.
Then, the selected clients optimize recommendation loss functions (e.g., E.q.~\ref{eq_ori_loss}) based on the received public parameters and their private parameters:
\begin{equation}\label{eq_ori_loss}
  \mathcal{L}^{rec} = -\sum\nolimits_{(u_{i}, v_{j}, r_{ij})\in \mathcal{D}_{i}} r_{ij}\log \hat{r}_{ij} + (1-r_{ij})\log (1-\hat{r}_{ij})
\end{equation}
After the local training, clients directly update their local private parameters while sending the updates of public parameters $\nabla \mathbf{V}^{t-1}$ and $\nabla \mathbf{\Theta}^{t-1}$ back to the central server.
The central server aggregates received gradients using FedAvg~\cite{mcmahan2017communication}.

\textbf{Base Recommender.}
Following most research~\cite{zhang2022pipattack,rong2022poisoning,yuan2023manipulating} of FedRecs' security, we employ neural collaborative filtering (NCF)~\cite{he2017neural} as the basic recommender model.
NCF is based on the matrix factorization.
It leverages multi-layer perceptron (MLP) and the concatenation of user and item embeddings to predict the user's preference score:
\begin{equation}\label{eq_ncf}
  \hat{r}_{ij} = \sigma(\mathbf{h}^{\top}MLP([\mathbf{u}_{i}, \mathbf{v}_{j}]))
\end{equation}
where $[\cdot]$ is a concatenation operation and $\mathbf{h}$ is trainable parameters.
It is worth noting that the above-mentioned general federated recommendation framework and our proposed methods are compatible with most deep learning-based recommenders~\cite{bai2017neural,he2018outer} since we do not make special assumptions on base recommenders.

\subsection{Model Poisoning Attacks in Federated Recommendation}
While the federated recommendation framework outlined in Section~\ref{sec_prim_fedrec} offers privacy protection for users, its open characteristics leave a backdoor for adversaries to directly alter the recommender model's public parameters via uploading poisoned gradients, namely model poisoning attacks~\cite{zhang2022pipattack}.
In this paper, we mainly focus on targeted model poisoning attacks~\cite{zhang2022pipattack,rong2022fedrecattack,yuan2023manipulating,rong2022poisoning}, which aim to promote target items $\widetilde{\mathcal{V}}$ with financial incentives, since these kind of attacks are more imperceptible and will cause unfair recommendation~\cite{rong2022fedrecattack}.
E.q.~\ref{eq_er} formally defines these model poisoning attacks, where $\hat{\mathcal{V}_{i}}$ is the recommendation result for user $u_{i}$. $\mathcal{V}_{i}^{-}$ is the set of items that $u_{i}$ has not interacted with.
$\nabla \widetilde{\mathbf{V}}^{t}$ and $\nabla \widetilde{\mathbf{\Theta}}^{t}$ are poisoned gradients uploaded by compromised clients.
$ER@K$ is the exposure ratio. It reflects the proportion of users for whom the target items appear in their top-K recommendation lists.
\begin{equation}
  \label{eq_er}
  \begin{aligned}
    &\mathop{argmax}\limits_{\{\nabla \widetilde{\mathbf{V}}^{t}, \nabla \widetilde{\mathbf{\Theta}}^{t}\}_{t=s}^{T-1}} ER@K (\widetilde{\mathcal{V}}|\mathbf{U}^{T}, \mathbf{V}^{T}, \mathbf{\Theta}^{T})\\
    ER@K &= \frac{1}{\left|\widetilde{\mathcal{V}} \right|} \sum\limits_{v_{j} \in \widetilde{\mathcal{V}}} \frac{\left|\left\{ u_{i} \in \mathcal{U} | v_{j}\in \hat{\mathcal{V}}_{i} \land v_{j}\in \mathcal{V}_{i}^{-} \right\}\right|}{\left|\left\{ u_{i} \in \mathcal{U} | v_{j}\in \mathcal{V}_{i}^{-} \right\} \right|}   
  \end{aligned}
\end{equation}
As of now,~\cite{rong2022poisoning,yuan2023manipulating} are two state-of-the-art model poisoning attacks that achieve high $ER@K$ scores with fewer assumptions, thereby exposing the vulnerability of FedRecs in scenarios that closely resemble real-world conditions. 
Therefore, we employ these two attacks to evaluate the robustness of FedRecs.

\section{Methodology}
This section presents our contrastive learning framework customized for federated recommender systems.
Specifically, we first introduce \modelname, consisting of user and item contrastive learning.
Subsequently, we empirically find that although \modelname improves the recommendation effectiveness, it exacerbates the vulnerability of FedRecs to model poisoning attacks.
We propose a popularity-based regularizer to address this security concern, resulting in a robust version of \modelname, denoted as \rmodelname. 
Algorithm~\ref{alg_fedrec} outlines \rmodelname in a generic FedRec through pseudocode.

\subsection{\modelname}
The main challenge of applying contrastive learning to FedRecs is constructing high-quality views within limited data $\mathcal{D}_{i}$ on each client's local device.
In the following parts, we present how \modelname constructs contrastive views for users and items respectively.

\subsubsection{User Contrastive Learning}\label{sec_user_contrastive_learning}
Given a user $u_{i}$, obtaining positive and negative samples for the user in FedRecs is challenging since the learning protocol strictly constrains that a client cannot access other clients' embeddings and private data.
\cite{wu2022fedcl,luo2023perfedrec++} radically overlook the protocol requirement to enable user embeddings shared so that the conventional contrastive learning methods can be simply applied in their works.
However, the exposure of user embeddings will cause severe privacy leakage as it violates the basic FedRec learning protocol~\cite{yuan2023interaction,zhang2023comprehensive}.
In light of this, we propose a privacy-preserving user contrastive learning method.

\textbf{Negative User Sample Construction.}
To facilitate clients construct negative users, the central server in \modelname manages a group of synthetic users $\mathcal{U}_{syn}$.
All the normal clients can access these synthetic users' private parameters and utilize them as negative users. 
Specifically, the central server selects a set of items for each synthetic user $s_{i}$ as its interacted items.
To align the data distribution of synthetic users with that of normal users (i.e., power-law distribution of interaction data), the items are randomly selected, taking into account item popularity:
\begin{equation}\label{eq_syn_data_select}
  \mathcal{D}_{syn,i} \leftarrow\mathop{random} (\mathcal{V}|popularity(\mathcal{V}), N)
\end{equation}
where $N$ is the number of interacted items for a synthetic user.
$popularity(\mathcal{V})$ is the popularity information of each item, which is one of the common commercial statistics available in many real-world applications (e.g., video view counts in YouTube\footnote{\url{https://www.youtube.com/}} and music listen counts in Spotify\footnote{\url{https://www.spotify.com/}}).

Then, the central server calculates synthetic user embedding by optimizing the recommendation loss function based on these synthetic data and the up-to-date public parameters:
\begin{equation}\label{eq_syn_user_emb}
  \mathbf{s}_{i}^{t} \leftarrow \mathop{argmin} \mathcal{L}^{rec}(\mathbf{s}_{i}^{t-1}|\mathbf{V}^{t-1},\mathbf{\Theta}^{t-1},\mathcal{D}_{syn,i})
\end{equation}
It is worth mentioning that E.q.~\ref{eq_syn_user_emb} only updates synthetic user embedding and will not influence the public parameters, as these synthetic users are developed to simulate negative user samples.
After that, these synthetic user embeddings $\{\mathbf{s}_{i}^{t}\}_{s_{i} \in \mathcal{U}_{syn}}$ are dispersed to normal clients to function as negative user embeddings.

\textbf{User View Augmentation for Positive Sample Construction.}
Inspired by~\cite{yu2022graph}, the normal client augments its user embedding $\mathbf{u}_{i}$ by adding noise vectors:
\begin{equation}\label{eq_user_aug}
  \begin{aligned}
    \mathbf{u}_{i}' = \mathbf{u}_{i} + \boldsymbol{\epsilon}_{i}',\quad \mathbf{u}_{i}'' = \mathbf{u}_{i} + \boldsymbol{\epsilon}_{i}''
  \end{aligned}
\end{equation}
We omit the time index to make the formula clear. 
$\boldsymbol{\epsilon}$ is based on the noise vector sampled from the uniform distribution $\bar{\boldsymbol{\epsilon}}\sim uniform(\mathbf{0},\mathbf{I})$ with the following constraints to avoid too much deviation of augmented representation: 
$\boldsymbol{\epsilon} = \bar{\boldsymbol{\epsilon}} \odot sign(\mathbf{u}_{i})$ and $\left\|\boldsymbol{\epsilon} \right\|_{2}=\eta$.

\textbf{User Contrastive Objective.}
Based on the synthetic user embeddings $\{\mathbf{s}_{i}^{t}\}_{s_{i} \in \mathcal{U}_{syn}}$ and augmented user views $\mathbf{u}_{i}'$ and $\mathbf{u}_{i}''$, the client $u_{i}$ optimizes the following loss function:
\begin{equation}\label{eq_user_con_loss}
\small
  \mathcal{L}^{uc} = -log \frac{exp(sim(\mathbf{u}_{i}',\mathbf{u}_{i}'')/\tau)}{exp(sim(\mathbf{u}_{i}',\mathbf{u}_{i}'')/\tau) + \sum_{s_{j}\in \mathcal{U}_{syn}} exp(sim(\mathbf{u}_{i},\mathbf{s}_{j})/\tau)}
\end{equation}
where $sim(x,y)$ is the similarity score of $x$ and $y$ calculated using cosine similarity and $\tau$ is the temperature hyper-parameter.
In E.q.~\ref{eq_user_con_loss}, only $\mathbf{u}_{i}$ is trainable while all synthetic user embeddings are frozen.

\subsubsection{Item Contrastive Learning}\label{sec_item_contrastive_learning}
Compared to users, the negative item samples are relatively straightforward as each client often has more than one training item in $\mathcal{D}_{i}$.
In this paper, for one training item, we simply treat the other items in the training set as negative items as these items refer to different entities.

\textbf{Item View Augmentation.}
A direct approach to augment item representation involves the application of E.q.~\ref{eq_user_aug} to item embeddings. However, our empirical findings, as outlined in Table~\ref{tb_view_construct_impacts}, indicate that the performance gains from this method are limited. This limitation may arise due to the scarcity of training data on each client side, preventing the model from effectively learning meaningful knowledge amidst the presence of relatively inconsequential uniform noise. In response to this challenge, we introduce a more potent technique by augmenting item views with more purposeful ``noise''. 
Specifically, based on the augmented user representations $\mathbf{u}_{i}'$ and $\mathbf{u}_{i}''$, \modelname 
can obtain the corresponding item embedding updates as follows:
\begin{equation}\label{eq_item_emb_noise}
\small
  \boldsymbol{\Delta}' \leftarrow \mathop{argmin} \mathcal{L}^{rec}(\mathbf{V}|\mathbf{u}_{i}',\mathbf{\Theta},\mathcal{D}_{i}),~  \boldsymbol{\Delta}'' \leftarrow \mathop{argmin} \mathcal{L}^{rec}(\mathbf{V}|\mathbf{u}_{i}'',\mathbf{\Theta},\mathcal{D}_{i})
\end{equation}
Note that since E.q.~\ref{eq_item_emb_noise} is only employed for item view augmentation, there is no need to meticulously compute updates over multiple epochs. In contrast, we opt to perform a single forward and backward pass to approximate the updates. This strategy is adopted to mitigate the burden of high computational costs.
Subsequently, the item view is augmented as follows:
\begin{equation}\label{eq_item_aug}
  \begin{aligned}
    \mathbf{V}_{i}' = \mathbf{V}_{i} + \boldsymbol{\Delta}_{i}',\quad \mathbf{V}_{i}'' = \mathbf{V}_{i} + \boldsymbol{\Delta}_{i}''
  \end{aligned}
\end{equation}

\textbf{Item Contrastive Objective.}
Based on the augmented item views, \modelname optimizes the following contrastive loss function:
\begin{equation}\label{eq_item_con_loss}
  \mathcal{L}^{ic} = \sum_{v_{k}\in\mathcal{D}_{i}} -log \frac{exp(sim(\mathbf{v}_{k}',\mathbf{v}_{k}'')/\tau)}{\sum_{v_{j}\in \mathcal{D}_{i}} exp(sim(\mathbf{v}_{k}',\mathbf{v}_{j}'')/\tau)}
\end{equation}

\subsubsection{Joint Learning with Recommendation Task}
To improve the recommendation performance, we jointly train the contrastive learning tasks (E.q.~\ref{eq_user_con_loss} and E.q.~\ref{eq_item_con_loss}) with the recommendation task (E.q.~\ref{eq_ori_loss}) on each client device:
\begin{equation}\label{eq_cl4fedrec_loss}
  \mathcal{L} = \mathcal{L}^{rec} + \lambda_{1} \mathcal{L}^{uc} + \lambda_{2} \mathcal{L}^{ic}
\end{equation}
where $\lambda_{1}$ and $\lambda_{2}$ are hyper-parameters to control the strengths of user contrastive learning and item contrastive learning, respectively.

\begin{figure*}[!htbp]
  \centering
  \includegraphics[width=1.\textwidth]{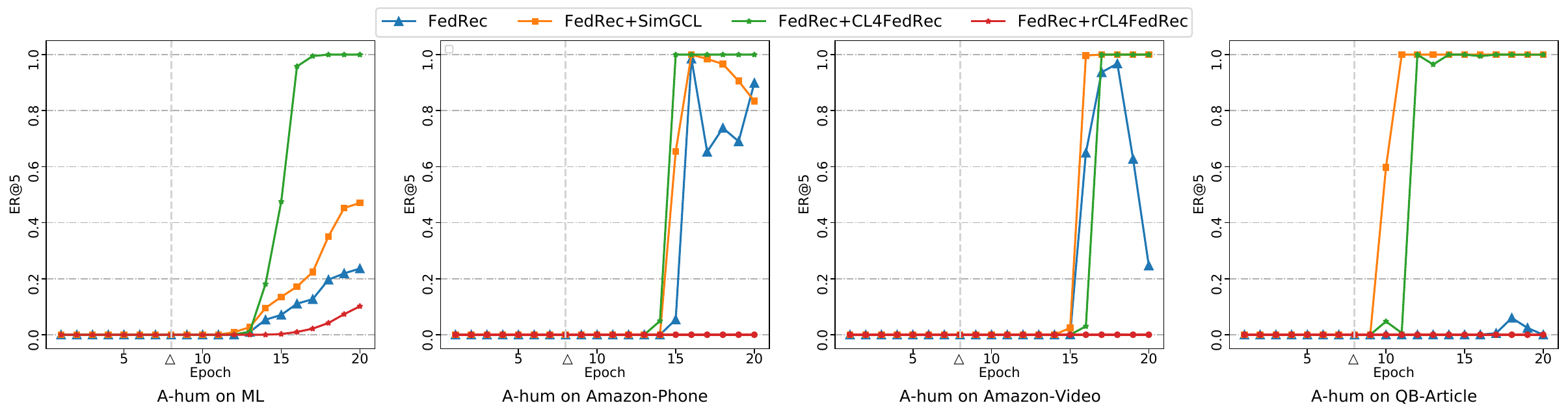}
  \caption{ER@5 scores of A-hum attack FedRec with and without contrastive learning methods on different recommendation datasets.}\label{fig_a-hum_for_attacks}
\end{figure*}
\begin{figure*}[!htbp]
  \centering
  \includegraphics[width=1.\textwidth]{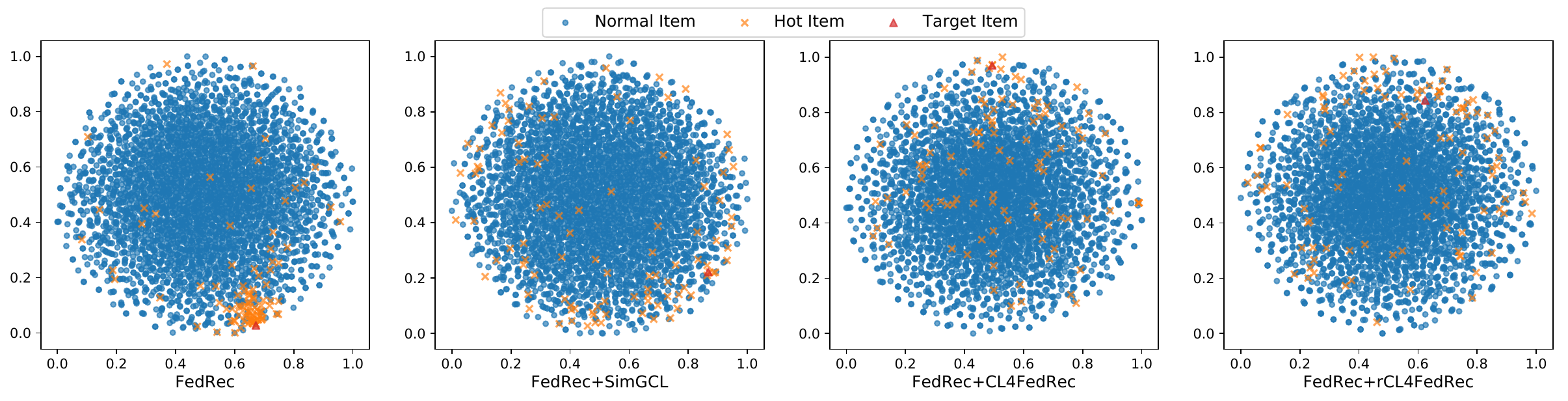}
  \caption{Item representation distribution on MovieLens-1M for FedRecs with and without different contrastive learning methods. We define the top 100 items that have the most number of interactions as ``hot item''.}\label{fig_emb_dist}
\end{figure*}
\begin{figure}[!htbp]
  \centering
  \subfloat[The FedRec with different strengths of $\mathcal{L}^{uni}$.]
  {\includegraphics[width=0.42\textwidth]{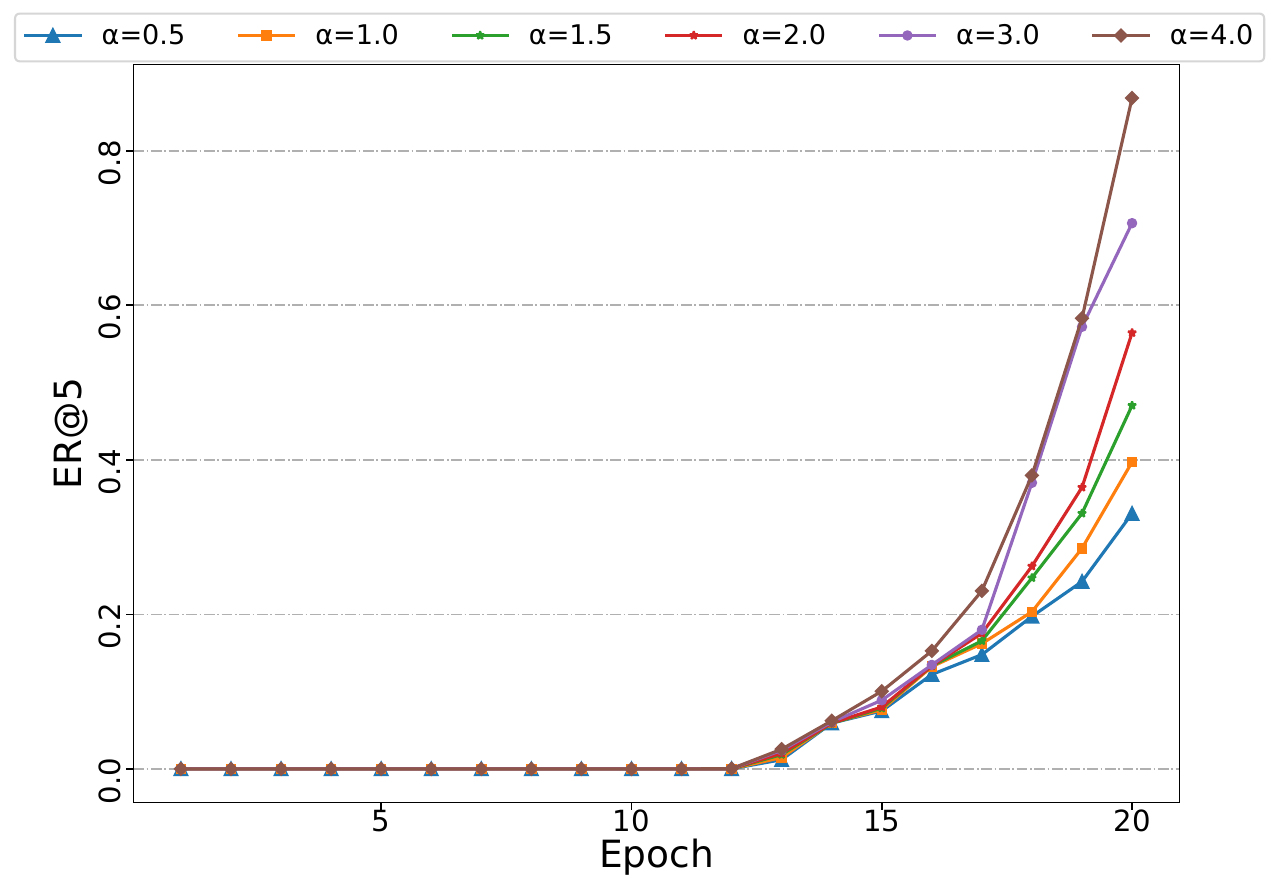}\label{fig_dim_attack}}
  \hfil
  \subfloat[The sorted singular value of item embedding table for FedRec with different strengths of $\mathcal{L}^{uni}$.]{\includegraphics[width=0.4\textwidth]{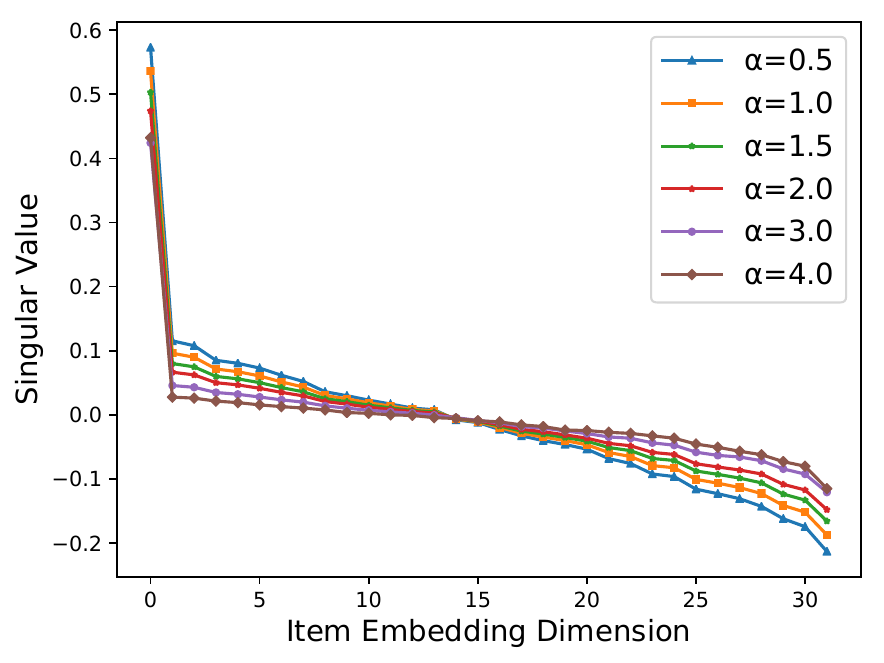}\label{fig_dim_singular}}
  \caption{Proof-of-concept. More uniform distribution (i.e., larger $\alpha$) weakens the robustness of FedRecs to model poisoning attacks.}\label{fig_dim_proof}
  \vspace{-15pt}
\end{figure}

\subsection{Robust \modelname against Model Poisoning Attacks}
The empirical results discussed in Section~\ref{sec_effectiveness_cl} demonstrate a noteworthy improvement of FedRecs in the recommendation performance brought by \modelname. As FedRecs have been shown to be susceptible to model poisoning attacks, we take a further step to assess the robustness of \modelname. 

\subsubsection{Robustness Problem}
We employ one of the state-of-the-art model poisoning attacks, A-hum~\cite{rong2022poisoning}, to attack FedRecs enhanced by various contrastive learning methods to highlight their vulnerability.
As illustrated in Fig.~\ref{fig_a-hum_for_attacks}, compared to the original FedRec, A-hum achieves higher and rapidly increasing ER@5 scores in FedRecs with contrastive learning methods across all four datasets. 
This observation suggests that \textbf{simply incorporating contrastive learning methods renders FedRecs more susceptible to model poisoning attacks}. In the experiment section, we will employ more state-of-the-art model poisoning attacks to assess the robustness of FedRecs with various contrastive learning methods.

\subsubsection{Emperical Analysis}
We posit that this phenomenon arises from the uniformity and dispersion of embedding distribution induced by contrastive learning~\cite{wang2020understanding,yu2022graph}.
Fig.~\ref{fig_emb_dist} plots different FedRecs' item embeddings trained on MovieLens-1M, and similar distribution can also be observed on the other three datasets.
In the original FedRec, most hot items' embeddings are densely clustered.
Consequently, if an adversary aims to elevate a target item's popularity, it must ``precisely'' shift the embedding of the target item closer to the cluster of popular items. 
However, when integrating a contrastive learning task, popular item embeddings are globally dispersed among normal item embeddings, as depicted in the three right figures in Fig.~\ref{fig_emb_dist}.
Since the embeddings of popular items are scattered in the embedding space, the adversary has more opportunities to manipulate the embedding of its target item to disguise it as a popular item.

To further validate our claim that the distribution uniformity makes FedRecs vulnerable, we provide a proof-of-concept in Fig.~\ref{fig_dim_proof}. 
Specifically, inspired by~\cite{hua2021feature,shi2022towards,yuan2023hetefedrec}, we employ the following formula to directly adjust the ``uniformity'' of item embeddings:
\begin{equation}\label{eq_uniformity_loss}
  \mathcal{L}^{uni} = \frac{1}{d}\left\| corr(\frac{\mathbf{V} - \bar{\mathbf{V}}}{\sqrt{var(\mathbf{V})}}) \right\|_{F}
\end{equation}
where $d$ is the dimension of item embeddings, $corr(\cdot)$ is the function that calculates the correlation matrix, $\bar{\mathbf{V}}$ is the mean of each dimension, $var(\cdot)$ computes the variance of a matrix, and $\left\|\cdot\right\|_{F}$ represents the Frobenius norm. 
In the proof-of-concept, the client optimizes $\mathcal{L}^{uni}$ with the recommendation loss function as follows:
\begin{equation}\label{eq_proof_loss}
  \mathcal{L}^{poc} = \mathcal{L}^{rec} + \alpha\mathcal{L}^{uni}
\end{equation}
where $\alpha$ controls the strength of the uniformity penalty.
As demonstrated in Fig.~\ref{fig_dim_singular}, with larger $\alpha$, the variance of each dimension's singular value is diminished, i.e., the curve becomes less steep, indicating a more uniform embedding table. Then, we employ A-hum to assess the system's vulnerability with varying $\alpha$.
As depicted in Fig.~\ref{fig_dim_attack}, it becomes evident that as $\alpha$ increases, A-hum achieves higher ER@5 scores.
This observation confirms our assertion that the uniformity of distribution weakens the resilience of FedRecs against model poisoning attacks.

\subsubsection{Solution: Popularity-based Contrastive Regularizer}
Based on the finding that the uniformity and dispersion of item embeddings enable adversaries to easily push the target item's representation close to popular items, we propose a popularity-based regularizer to calibrate item embeddings.
In detail, the central server divides the items into hot items and normal items according to the popularity information.
Then, at the end of each global epoch, the central server updates the item embeddings by optimizing the following contrastive loss function:
\begin{equation}\label{eq_regularize_loss}
  \mathcal{L}^{reg} = \sum_{v_{i}\in\mathcal{V}^{hot}} -log \frac{exp(1/\tau)}{\sum_{v_{j}\in \mathcal{V}^{sp}} exp(sim(\mathbf{v}_{i},\mathbf{v}_{j})/\tau)}
\end{equation}
where $\mathcal{V}^{hot}$ is the set of hot items and $\mathcal{V}^{sp}$ is the subset of items sampled from the normal item set.
E.q.~\ref{eq_regularize_loss} pulls the normal items away from hot items. Therefore, malicious users cannot easily push their target items to mimic popular items.

\begin{algorithm}[!ht]
  \renewcommand{\algorithmicrequire}{\textbf{Input:}}
  \renewcommand{\algorithmicensure}{\textbf{Output:}}
  \caption{FedRec with \rmodelname.} \label{alg_fedrec}
  \begin{algorithmic}[1]
    \Require global epoch $T$, local epoch $L$, learning rate $lr$, client batch size $B$, \dots
    \Ensure public parameters $\mathbf{V}, \mathbf{\Theta}$, private parameters $\mathbf{u}_{i}|_{i \in \mathcal{U}}$
    \State initialize global parameter $\mathbf{V}_{0}, \mathbf{\Theta}_{0}$
    \State construct synthetic users $\mathcal{U}_{syn}$ using E.q.~\ref{eq_syn_data_select}
    \For {training round t = 1, ..., $T$}
      \State shuffle and divide $\mathcal{U}$ into $\{\mathcal{U}_{k}\}_{k=1\dots\left|\mathcal{U}\right|/B}$
      \For{k=1\dots$\left|\mathcal{U}\right|/B$}
        \For{$u_{i}\in \mathcal{U}_{k}$ \textbf{in parallel}}
          \State \Call{ClientTrain}{$\{\mathbf{s}_{j}^{t}\}_{s_{j} \in \mathcal{U}_{syn}},\mathbf{V}^{t-1}, \mathbf{\Theta}^{t-1}$}
        \EndFor
        \State $\mathbf{V}^{t-1}, \mathbf{\Theta}^{t-1}\leftarrow$ aggregate public parameter updates
      \EndFor
      \State $\mathbf{V}^{t} \leftarrow$ calibrate item embedding using E.q.~\ref{eq_regularize_loss}
      \State $\{\mathbf{s}_{j}^{t+1}\}_{s_{j} \in \mathcal{U}_{syn}}\leftarrow$update synthetic users using E.q.~\ref{eq_syn_user_emb}.
    \EndFor
    \Function{ClientTrain} {$\{\mathbf{s}_{j}^{t}\}_{s_{j} \in \mathcal{U}_{syn}},\mathbf{V}^{t-1}, \mathbf{\Theta}^{t-1}$}
      \State augment user views using E.q.~\ref{eq_user_aug}
      \State augment item views using E.q.~\ref{eq_item_emb_noise},\ref{eq_item_aug}
      \State $\nabla \mathbf{V}^{t-1}_{i}, \nabla \mathbf{\Theta}^{t-1}_{i}, \nabla \mathbf{u}^{t-1}_{i}\leftarrow$ train $L$ epochs on $\mathcal{D}_{i}$ using E.q.~\ref{eq_cl4fedrec_loss}
      \State $\mathbf{u}_{i}^{t} \leftarrow$ update private parameter using $\nabla \mathbf{u}^{t-1}_{i}$ 
      \State upload $\nabla \mathbf{V}^{t-1}_{i}, \nabla \mathbf{\Theta}^{t-1}_{i}$
    \EndFunction
    \end{algorithmic}
\end{algorithm}

\begin{table}[!htbp]
  \caption{Statistics of recommendation datasets.}\label{tb_dataset_statistics}
  \resizebox{0.47\textwidth}{!}{
  \begin{tabular}{l|ccccc}
  \hline
  \textbf{Dataset}      & \textbf{\#users} & \textbf{\#items} & \textbf{\#interactions} & \textbf{Avg.} & \textbf{Sparsity} \\ \hline
  \textbf{MovieLens-1M} & 6,040          & 3,706          & 1,000,208             & 165.5         & 95.53\%           \\
  \textbf{Amazon-Phone} & 13,174         & 5,970          & 103,593               & 7.8           & 99.86\%           \\
  \textbf{Amazon-Video} & 8,072          & 11,830         & 63,836                & 7.9           & 99.93\%           \\
  \textbf{QB-Article}   & 10,981         & 6,493          & 335,663               & 30.5          & 99.52\%           \\ \hline
  \end{tabular}}
\end{table}

\section{Experiments}
In this section, we conduct extensive experiments to answer the following research questions (RQs):
\begin{itemize}
  \item \textbf{RQ1.} How effective are our proposed federated contrastive learning methods (\modelname and \rmodelname) to improve the recommendation performance?
  \item \textbf{RQ2.} How robust are our proposed federated contrastive learning methods (\modelname and \rmodelname) to state-of-the-art model poisoning attacks?  As a new defense method, how effective is our proposed popularity-based contrastive regularizer in improving the robustness of FedRecs compared to existing defense baselines?
  \item \textbf{RQ3.} What is the contribution of different components in \rmodelname to the recommendation performance? 
  \item \textbf{RQ4.} What is the impact of hyper-parameters to \rmodelname?
\end{itemize}

\begin{table*}[!htbp]
  \centering
  \caption{Recommendation effectiveness comparison of various contrastive learning methods on four recommendation datasets.}\label{tb_main_result}
  \begin{tabular}{l|cc|cc|cc|cc}
  \hline
  \multirow{2}{*}{\textbf{Methods}} & \multicolumn{2}{c|}{\textbf{MovieLens-1M}} & \multicolumn{2}{c|}{\textbf{Amazon-Phone}} & \multicolumn{2}{c|}{\textbf{Amazon-Video}} & \multicolumn{2}{c}{\textbf{QB-Article}} \\ \cline{2-9} 
                                    & \textbf{Recall@20}   & \textbf{NDCG@20}    & \textbf{Recall@20}   & \textbf{NDCG@20}    & \textbf{Recall@20}   & \textbf{NDCG@20}    & \textbf{Recall@20} & \textbf{NDCG@20}   \\ \hline
  \textbf{Original}                 & 0.04444            & 0.06372           & 0.05909            & 0.02571           & 0.04626            & 0.01757           & 0.06244          & 0.03345          \\
  \textbf{SimGCL}                   & 0.04619            & 0.06252           & 0.05962            & 0.02601           & 0.04897            & 0.01994           & 0.06838          & 0.03512          \\
  \textbf{UNION}                    & 0.03334            & 0.05606           & 0.05860            & 0.02637           & 0.04954            & 0.02059           & 0.06105          & 0.03582          \\ \hline
  \textbf{\modelname}               & 0.04656            & 0.06842           & \textbf{0.05974}   & 0.02654           & 0.05350            & 0.02011           & \textbf{0.07413} & 0.03915          \\
  \textbf{\rmodelname}              & \textbf{0.04836}   & \textbf{0.07103}  & 0.05932            & \textbf{0.02782}  & \textbf{0.05667}   & \textbf{0.02117}  & 0.06962          & \textbf{0.04647} \\ \hline
  \end{tabular}
  \end{table*}

\subsection{Datasets}
We conduct experiments on four popular recommendation datasets: MovieLens-1M~\cite{harper2015movielens}, Amazon-Phone, Amazon-Video~\cite{mcauley2015image}, and QB-Article~\cite{yuan2022tenrec}, covering various platforms and recommendation domains.
MovieLens-1M includes $1,000,208$ interaction records between $6,040$ users and $3,706$ movies. 
Amazon-Phone has $13,174$ users, $5,970$ cell phones, and $103,593$ feedbacks.
There are $63,836$ interactions involving $8,072$ users and $11,830$ items in Amazon-Video.
QB-Article contains $10,981$ users, $6,493$ articles, and $335,663$ reading records.
Following the common settings in implicit feedback recommendation~\cite{rong2022poisoning,yuan2023manipulating,he2017neural}, all users' feedback ratings are transformed to $r_{ij}=1$, and negative instances are sampled with $1:4$ ratio.
Besides, we leave $20\%$ data for testing, and $10\%$ data are sampled from training data for validation.

\subsection{Evaluation Protocol}
In this paper, we employ two widely used metrics~\cite{yu2022graph} Recall at rank 20 (Recall@20) and Normalized Discounted Cumulative Gain at rank 20 (NDCG@20) to measure the recommendation effectiveness.
Recall reflects the average probability of ground-truth items successfully appearing in users' recommendation lists, while NDCG considers the position of ground-truth items. We rank all items when calculating the two metrics.

To evaluate the robustness of FedRecs, we primarily employ two state-of-the-art model poisoning attacks, selected for their ability to achieve remarkable performance without requiring extensive prior knowledge:
\begin{itemize}
  \item \textbf{A-hum}~\cite{rong2022poisoning}: This method utilizes ``hard users'' who consider the target items as negative samples to generate poisoned gradients to optimize E.q.~\ref{eq_er}.
  \item \textbf{PSMU}~\cite{yuan2023manipulating}: This method leverages randomly constructed users to generate poisoned gradients and further improve target items' competition by adding their alternative items to optimize E.q.~\ref{eq_er}.
\end{itemize}
In line with the original papers describing these attacks~\cite{rong2022poisoning, yuan2023manipulating}, we use the Exposure Ratio at rank 5 (ER@5) to quantify the effectiveness of the attacks. Higher and rapidly increasing ER@5 scores indicate more potent attacks and, consequently, reduced robustness of the FedRecs.

\subsection{Baselines}
\textbf{Contrastive Learning Baselines.}
As discussed in Section~\ref{sec_related_cl_in_rs}, most contrastive learning methods in centralized recommender models cannot work in FedRecs due to the extreme sparsity of local data on a client.  
Besides, as we focus on applying contrastive learning without compromising user privacy, those federated contrastive learning works~\cite{wu2022fedcl,luo2023perfedrec++} that need to share user embeddings are not considered in our baselines for fair comparison.
\begin{itemize}
  \item \textbf{Original}: This method shows the original state (i.e., recommendation effectiveness and robustness) of FedRecs without using any contrastive learning.
  \item \textbf{SimGCL}~\cite{yu2022graph}: This is the state-of-the-art contrastive learning method for centralized recommender systems, and it can directly augment the user/item embeddings by adding random noise. 
  \item \textbf{UNION}~\cite{yu2023untargeted}: This is the only contrastive learning method in FedRecs that does not sacrifice user privacy. It naively treats clients' all interacted items as positive samples and non-interacted items as negative samples to build item views, and it only considers the item contrastive learning.
\end{itemize}

\begin{figure*}[]
  \centering
  \subfloat[Regularizer and defense baselines against PSMU.]{\includegraphics[width=1\textwidth]{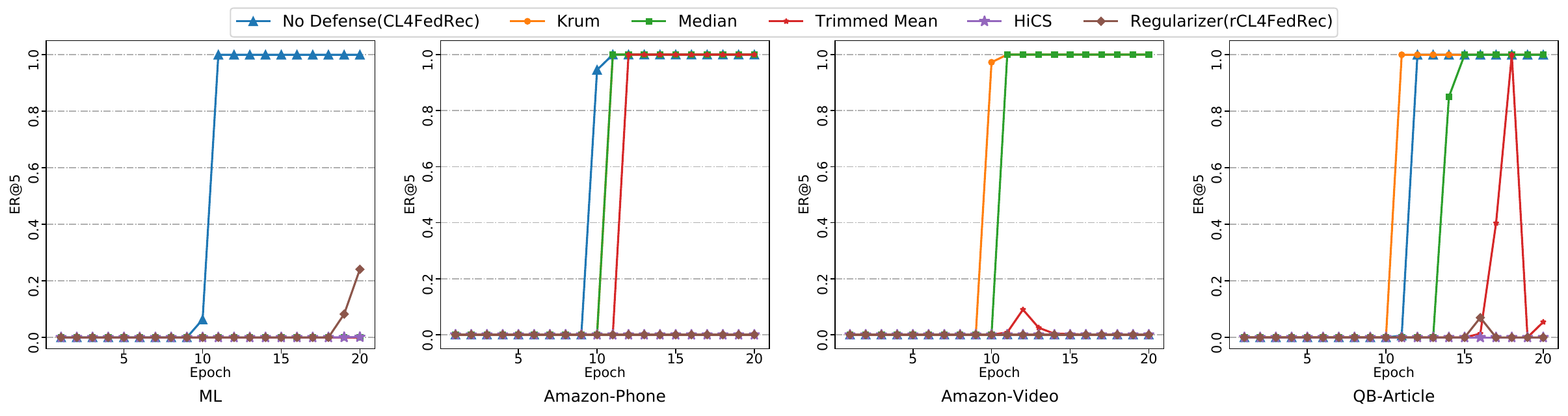}\label{fig_regularize_against_psmu}}
  \hfil
  \subfloat[Regularizer and defense baselines against A-hum.]{\includegraphics[width=1\textwidth]{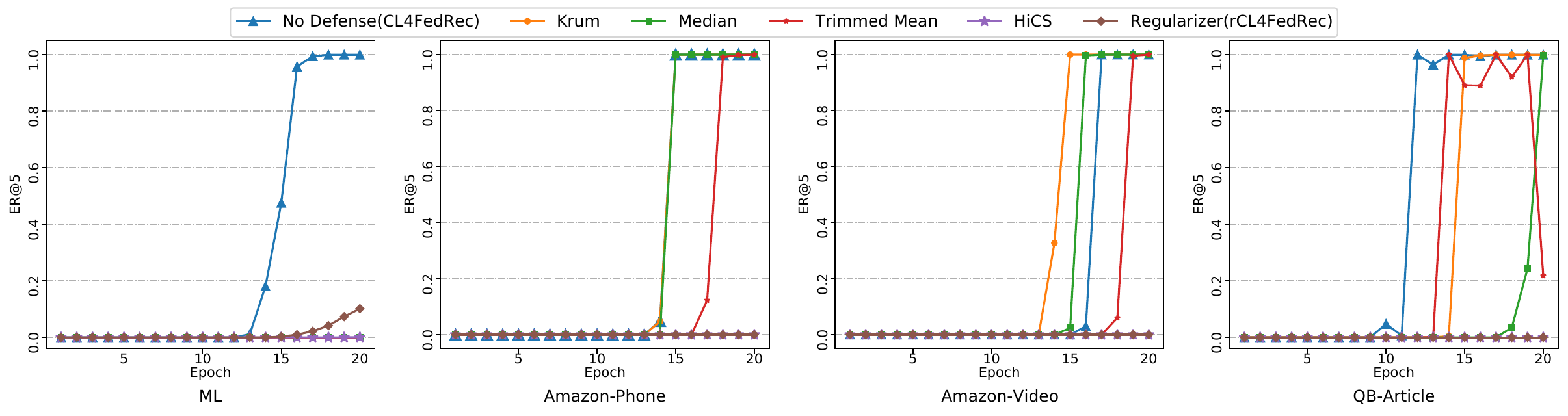}\label{fig_regularize_against_ahum}}
  \caption{The comparison of our regularizer with other defense baselines against state-of-the-art attacks.}\label{fig_regularizer_against_attacks}
\end{figure*}

\textbf{Defense Baselines.} We compare our proposed defense method, the popularity-based regularizer, with the following defense baselines by integrating them into \modelname.
\begin{itemize}
  \item \textbf{Krum}~\cite{blanchard2017machine}: We mainly apply Krum to the public parameters (e.g., item embeddings) on the server, i.e., we select the gradient of an item that is closest to the mean of all other clients' uploaded gradients of this item as the aggregated gradient.
  \item \textbf{Median}~\cite{yin2018byzantine}: This method chooses the median value of sorted gradients as the aggregated gradients.
  \item \textbf{Trimmed Mean}~\cite{yin2018byzantine}: This method removes a parameter's largest and smallest gradients and aggregates the remaining gradients.
  \item \textbf{HiCS}~\cite{yuan2023manipulating}: This is a gradient-clipping-based defense tailored for FedRecs. It adopts gradient clipping with sparsification updates to limit the contributions of malicious users.
\end{itemize}

\subsection{Implementation Details}
The user and item embedding sizes in FedRecs are set to $32$. 
Three feedforward layers with dimensions $64, 32, 16$ are used to process the concatenated user and item embeddings.
Adam~\cite{kingma2014adam} with $0.001$ learning rate is adopted as the optimizer.
For \modelname, the synthetic users' interacted item size  $N$ is set to $30$.
$\left|\mathcal{U}_{syn}\right|$ is $60$ for MovieLens-1M and QB-Article while $10$ for Amazon-Phone and Amazon-Video, respectively.
$\eta$ is $0.1$, $\tau$ equals $0.2$, and $\lambda_{1}=\lambda_{2}=0.5$.
All the FedRecs are converged within $20$ global epochs.
All the settings of model poisoning attacks follow the original paper~\cite{yuan2023manipulating,rong2022poisoning} with $0.1\%$ malicious users. In addition, we also perform the sensitivity analysis of key hyper-parameters in Section~\ref{sec_hyper_analysis}.

\subsection{Recommendation Effectiveness (RQ1)}\label{sec_effectiveness_cl}
Table~\ref{tb_main_result} presents a comparison of the recommendation performance between our proposed methods and contrastive baselines.
Specifically, when incorporating UNION, the performance of FedRec either remains unchanged or even declines on most datasets. 
This is attributed to the overly simplistic item views in UNION and the lack of consideration for user contrastive learning.
Although SimGCL yields some improvements for FedRec, the gains are limited as the item view augmentation in SimGCL introduces only meaningless uniform noises.
Notably, the performance of \modelname surpasses all baselines across all datasets, as evident from both Recall and NDCG scores, underscoring the superiority of our contrastive learning framework in FedRecs.
Furthermore, the integration of the popularity-based contrastive regularizer in \modelname (i.e., \rmodelname) results in further performance improvement. 
In summary, our proposed methods, including \modelname and \rmodelname, effectively enhance the original FedRec's performance.

\begin{table*}[!htbp]
  \centering
  \caption{The impact of different defenses on recommendation performance.}\label{tb_defense_on_performance}
  \begin{tabular}{l|cc|cc|cc|cc}
    \hline
    \multirow{2}{*}{\textbf{Methods}} & \multicolumn{2}{c|}{\textbf{MovieLens-1M}} & \multicolumn{2}{c|}{\textbf{Amazon-Phone}} & \multicolumn{2}{c|}{\textbf{Amazon-Video}} & \multicolumn{2}{c}{\textbf{QB-Article}} \\ \cline{2-9} 
                                      & \textbf{Recall@20}    & \textbf{NDCG@20}   & \textbf{Recall@20}    & \textbf{NDCG@20}   & \textbf{Recall@20}    & \textbf{NDCG@20}   & \textbf{Recall@20}  & \textbf{NDCG@20}  \\ \hline
    \textbf{\modelname}                     & 0.04656               & 0.06842            & \textbf{0.05974}      & 0.02654            & 0.05350               & 0.02011            & \textbf{0.07413}    & 0.03915           \\ \hline
    \textbf{+Krum}                     & 0.02794               & 0.05004            & 0.03821               & 0.01600            & 0.03298               & 0.01330            & 0.04901             & 0.02468           \\
    \textbf{+Median}                   & 0.02772               & 0.05062            & 0.03710               & 0.01574            & 0.03095               & 0.01150            & 0.0476              & 0.02542           \\
    \textbf{+Trimmed Mean}             & 0.04308               & 0.06743            & 0.05695               & 0.02460            & 0.05222               & 0.01799            & 0.06527             & 0.03903           \\
    \textbf{+HiCS}                     & 0.04437               & 0.06984            & 0.05908               & 0.02775            & 0.05317               & 0.01990            & 0.06834             & 0.03847           \\ \hline
    \textbf{+Regularizer}              & \textbf{0.04836}      & \textbf{0.07103}   & 0.05932               & \textbf{0.02782}   & \textbf{0.05667}      & \textbf{0.02117}   & 0.06962             & \textbf{0.04647}  \\ \hline
    \end{tabular}
  \end{table*}

\subsection{Recommendation Robustness  (RQ2)}
One of the contributions in this paper is the popularity-based contrastive regularizer, designed to enhance the robustness of \modelname. 
In this section, we compare our regularizer with some defense baselines. 
In general, an effective defense plugin should meet the following two requirements: (1) It can reduce the performance of attacks; (2) It does not compromise model performance.

In Fig.~\ref{fig_regularizer_against_attacks}, we execute two state-of-the-art attacks (A-hum~\cite{rong2022poisoning} and PSMU~\cite{yuan2023manipulating}) for \modelname, and then, we utilize several commonly used defenses to against these attacks.
For these two attacks, only our regularizer and the state-of-the-art defense baseline HiCS can successfully keep the target items' ER@5 scores at zero in most cases. 
Other baselines cannot consistently protect CL4FedRec on various datasets.

In Table~\ref{tb_defense_on_performance}, we further investigate the effects of various defenses on recommendation performance.
According to the results, Krum, Median, and Trimmed Mean severely compromise the recommender model, reducing these methods' utility.
HiCS slightly diminishes model performance in some cases (e.g., Amazon-Video and QB-Article). 
Our regularizer is the only defense that significantly improves the model's performance.

Overall, our proposed regularizer achieves comparable defense effects with the state-of-the-art defense HiCS, and our regularize is the only defense method that can further enhance model performance, demonstrating its effectiveness.

\begin{table}[]
  \centering
  \caption{The impact of regularizer, user contrastive learning, and item contrastive learning on recommendation performance on MovieLens-1M.}\label{tb_component_impact}
  \begin{tabular}{l|cc}
  \hline
  \textbf{Methods}      & \textbf{Recall@20} & \textbf{NDCG@20} \\ \hline
  \textbf{\rmodelname}         & \textbf{0.04836}   & \textbf{0.07103} \\ \hline
  \textbf{-regularizer} & 0.04656            & 0.06842          \\
  \textbf{-user contrastive learning}         & 0.04793                  & 0.06923                \\
  \textbf{-item contrastive learning}          & 0.04637                  & 0.06656               \\ \hline
  \end{tabular}
  \end{table}

  \begin{table}[]
    \centering
    \caption{The impact of different user and item view construction on recommendation performance on MovieLens-1M.}\label{tb_view_construct_impacts}
    \begin{tabular}{l|cc}
    \hline
    \textbf{Methods} & \textbf{Recall@20} & \textbf{NDCG@20} \\ \hline
    \textbf{\rmodelname}    & \textbf{0.04836}   & \textbf{0.07103} \\ \hline
    \textbf{random synthetic users}     & 0.04629                  & 0.06902                \\
    \textbf{SimGCL-based item views}     & 0.04679                  & 0.06365                \\ \hline
    \end{tabular}
    \end{table}

\subsection{Ablation Study (RQ3)}
In this section, we assess the contributions of different components to recommendation performance in \rmodelname.
\rmodelname comprises three main components: user contrastive learning, item contrastive learning, and the popularity-based contrastive regularizer.
In Table~\ref{tb_component_impact}, we individually remove these three components to demonstrate their impacts.
As we can see, removing any one of the components results in a performance drop, indicating that all components contribute to the enhanced performance.
Specifically, when removing the regularizer or item contrastive learning, the model's performance decreases from 0.048 to around 0.046 Recall@20 scores, while eliminating user contrastive learning causes about a 0.005 Recall@20 score drop. 
Due to space limitations, we only present the results on MovieLens-1M, but similar conclusions can also be observed on the other three datasets.

Besides, \rmodelname employs a popularity-based synthetic negative user construction method (i.e., E.q.~\ref{eq_syn_data_select} and~\ref{eq_syn_user_emb}) and an approximately optimization-based item view augmentation (i.e., E.q.~\ref{eq_item_emb_noise} and~\ref{eq_item_aug}) for user and item contrastive learning.
Therefore, we investigate the impacts of these two methods in Table~\ref{tb_view_construct_impacts}.
Specifically, we replace the popularity-based synthetic users with randomly constructed users and utilize SimGCL to replace our item view augmentation, respectively.
According to Table~\ref{tb_view_construct_impacts},  using these naive view construction methods cannot achieve comparable performance to \rmodelname, indicating the effectiveness of our user and item view construction methods.

\begin{figure}[!htbp]
  \centering
  \includegraphics[width=0.48\textwidth]{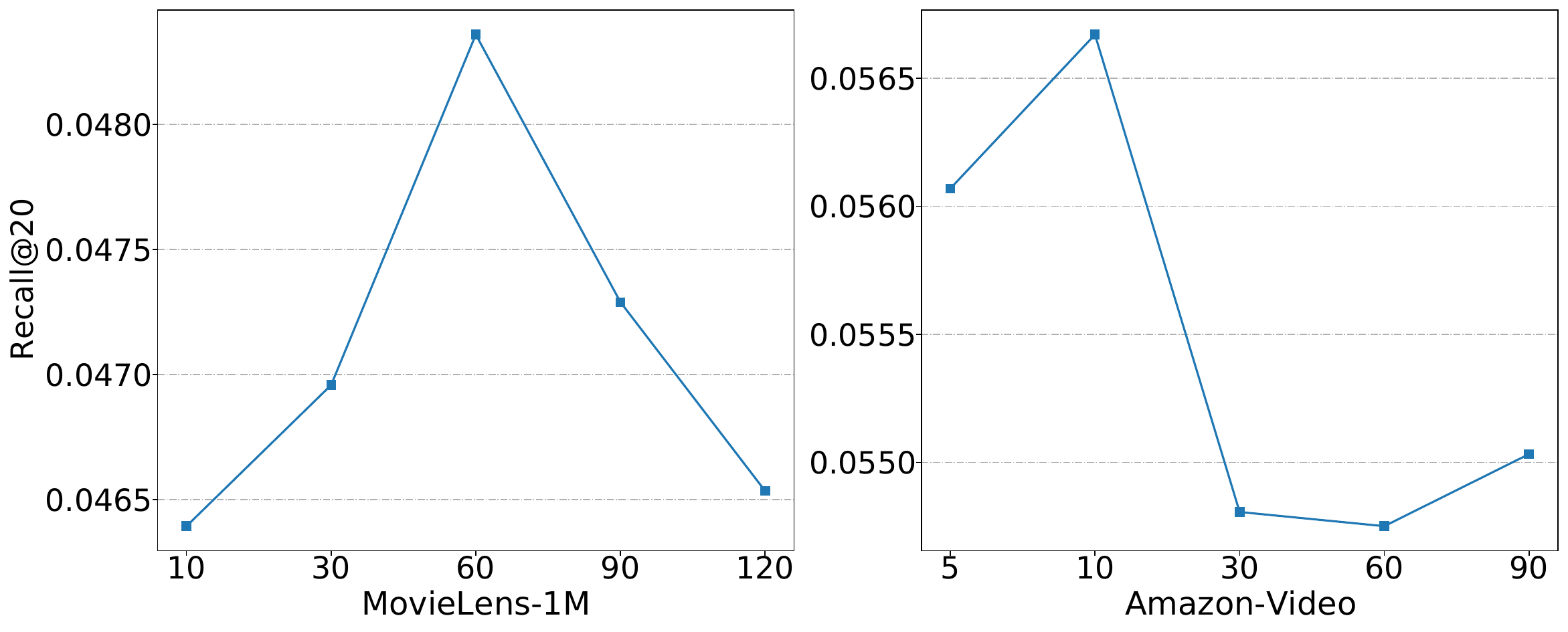}
  \caption{The performance trend w.r.t. the change of synthetic user numbers $\left|\mathcal{U}_{syn}\right|$ on MovieLens-1M and Amazon-Video.}\label{fig_hyperparameter}
\end{figure}

\subsection{Hyper-parameter Analysis (RQ4)}\label{sec_hyper_analysis}
The size of synthetic users (i.e., $\left|\mathcal{U}_{syn}\right|$) constructed for contrastive learning is a crucial hyper-parameter significantly influencing model performance. Fig.~\ref{fig_hyperparameter} shows the performance trends concerning $\left|\mathcal{U}_{syn}\right|$. 
According to the results, on MovieLens-1M, the model performance exhibited a positive correlation with the number of synthetic users until it reached 60. Beyond this threshold, the performance started to decrease. This is attributed to the fact that with fewer synthetic users, normal users struggle to learn from the negative users, while the synthetic users are too many, the contrastive learning is overwhelmed, impeding users learn recommendation knowledge.
On Amazon-Video, the optimal recommendation performance is achieved when $\left|\mathcal{U}_{syn}\right|$ is 10.
Note that due to space limitations, we only present the results on MovieLens-1M and Amazon-Video. The trend on QB-Article mirrors that of MovieLens-1M, and the trend on Amazon-Phone is similar to that on Amazon-Video.

\section{Conclusion}
In this paper, we introduce a contrastive learning framework tailored for federated recommender systems, namely \modelname, which designs user contrastive views by constructing synthetic users and constructs item contrastive views via approximate optimization.
Subsequently, we empirically observe that incorporating contrastive learning reduces the robustness of FedRecs under model poisoning attacks.
We attribute this phenomenon to the uniformity of the item embedding distribution. 
To address this, we propose a popularity-based contrastive regularizer for \modelname, forming the robust version (\rmodelname). Extensive experiments conducted on four datasets demonstrate the effectiveness and robustness of our proposed methods.

\begin{acks}
To Robert, for the bagels and explaining CMYK and color spaces.
\end{acks}

\bibliographystyle{ACM-Reference-Format}
\bibliography{sample-base}










\end{document}